# Microscopic characterisation of laser-written phenomena for component-wise testing of photonic integrated circuits

Jun Guan,[1,][*] Xiang Liu,[1] Adrian J. Menssen[2] and Martin J. Booth[1,3,][*]

[1] Department of Engineering Science, University of Oxford, Parks Road, Oxford, OX1 3PJ, UK

[2] Clarendon Laboratory, Department of Physics, University of Oxford, Oxford OX1 3PU, UK

[3] Centre for Neural Circuits and Behaviour, University of Oxford, Mansfield Road, Oxford, OX1 3SR, UK

Email: jun.guan@eng.ox.ac.uk; martin.booth@eng.ox.ac.uk

**Photonic integrated circuits (PICs) directly written with a femtosecond laser have shown great potential in many areas such as quantum information processing (QIP). Many applications, like photon-based quantum computing, demand the up-scaling of PICs and ever-higher optical performance, such as controllable polarisation dependence and lower loss. In order to overcome current limitations in fabrication precision, repeatability and material uniformity, a solution for non-destructive testing of large-scale PICs in a component-wise manner is desired to meet those ever-stricter demands. Here we demonstrate a solution for non-destructive component-wise testing by predicting the performance of a PIC component based on imaging with an adaptive optical third-harmonic-generation (THG) three-dimensional (3D) microscope. The 3D THG imaging can be performed on any component or part of it inside multi-component PIC. Moreover, through discovering new phenomena we also demonstrated that 3D THG microscopy provides a new pathway towards studying the fundamentals of light-matter interaction in transparent materials.**

By virtue of its rapid prototyping nature and flexibility with materials, femtosecond laser direct-writing has become a popular PIC fabrication technology for photon-based QIPs[1-8] among other application areas[9-14]. In particular, its intrinsic 3D fabrication capability uniquely enables fabrication of PICs involving arbitrary 3D photonic structures[15-18]. Photon-based QIPs impose strict requirements on the properties of PICs and components thereof in terms of loss[19], control of polarization[2] and other optical properties; those requirements will get ever stricter as the complexity of PICs increases to meet the demands of advanced QIPs like photon-based quantum computing. Those ever-stricter demands pose a great challenge to all the current PIC fabrication technologies[19-21], especially direct laser writing[19, 20] that faces fabrication errors, repeatability and substrate uniformity issues. The conventional light-coupling test, which is currently the only available way to test the overall performance of a PIC chip, is only a chip-wise test and cannot be used to individually diagnose a component or part of it inside a multicomponent PIC chip. Under those circumstances, being able to test components within large-scale PICs is vital to scaling up PICs, since the causes of underperformance of fabricated PICs can only be identified through component-wise testing. Once the causes, which can be fabrication errors of certain components or the PIC designs themselves, are identified, it is possible to improve the performance of PICs through guided iterative fabrication processes or improved designs, eventually to enable scaling up PICs. Bruck et al.[22] used a 'pump-probe' technique to spectrally test components of planar silicon PICs, but that technique does not meet the demand for component-wise testing of PIC in many areas like QIP, since that technique cannot be used to test the loss, polarization and many other non-spectral properties of PIC components.

Here, we demonstrate a method based on adaptive 3D THG microscopy to test the performance of a PIC in a component-wise manner, through ascertaining qualitative or quantitative links between the THG image of a PIC component and its properties, like loss, polarization and intensity splitting ratio, bearing in mind that any arbitrary location in the volume of a PIC chip can be non-destructively imaged in 3D with the THG microscope. Moreover, through discovering new phenomena, we demonstrate that this method also opens us a new pathway to studying the fundamentals of light-matter interaction in transparent materials.

In the following, we will first show the new pathway opened by adaptive THG microscopy through revealing new phenomena in femtosecond laser direct writing processes. Then the link between waveguide THG image and its loss, polarization properties and near field mode profile will be disclosed. Finally we will take the directional coupler and the three-waveguide coupler (tritter) as examples of complex PIC components to demonstrate how one can determine their performance based on their THG images.

**Effects of writing parameters on waveguide micro-morphology**

Many factors – such as wavelength, repetition rate, translation speed, pulse energy and duration – can affect the nature of laser written waveguides, leading to a complex parameter space within which one needs to identify an operating "sweet spot". The ability of THG microscopy to reveal structures in the cross-section of the waveguide provides an excellent opportunity to improve understanding of the processes that lead to optimal waveguide performance. Furthermore, this understanding can aid improved design of later generation PICs.

For our demonstrations, waveguides were written inside a borosilicate glass chip with different writing pulse energies and scan speeds (see Methods). THG images of their cross-sections were taken with a custom-built adaptive THG microscope (See Supplementary Section A), as shown in Fig. 1a. From Fig. 1a, we can see how the waveguide cross-sectional THG profiles evolved along with writing parameters. We note here that while the THG depends upon the third-order non-linear optical properties of the material, in practice the measured THG signal correlates spatially to changes in refractive-index[23, 24] profiles, such that a rapid spatial transition between different refractive indices typically correlates with a high signal. First of all, under the same pulse energy, the THG profile became more complex with decreasing scan speed, which indicated more complex refractive index changes at lower scan speeds. This is contradictory to the currently well-accepted purely-thermal effect[25]-based theory of photonic structure formation with high repetition rate femtosecond lasers, since a decrease of scan speed under constant pulse energy means increase of heating effects, which by intuition should result in smoother refractive index changes and simpler THG profile. Secondly, under the same scan speed, as pulse energy was increased, areas of stronger THG signal shifted from the top to the bottom of the cross-sectional profile. The detailed evolution of the THG profile along pulse energy was more complex. These phenomena cannot jointly be explained with existing theories – neither the well-accepted thermal theory[25] or less-established ones like ion migration[26]. One of the striking phenomena was the sudden emergence of much stronger THG signal at the bottom middle of THG profiles 4-7, 11-14, 20-21, 27-28 and 35. As exemplified in Fig. 1b, these areas corresponded to the dark spots at the bottom of the corresponding conventional white-light microscopy images, which were aligned with the focal plane location of writing objective lens[25]; although there was no obvious difference between those dark spots on white-light microscopy images of waveguide 19-20 and 26-27, the contrasts in terms of THG signal strength on their THG images at corresponding locations were striking. This suggests that there were actually two types of low-density region: a type I low-density region was formed through non-explosive rarefaction, for instance the marked regions on images of waveguide 19 and 26 in Fig. 1b; a type II low-density region was the broadly recognised one which is formed by micro-explosion[27, 28] and represented by the marked regions on the images of waveguide 20 and 27. The slight changes of writing pulse energy from waveguide 19 and 26 to waveguide 20 and 27 respectively implies that there was a threshold-like sharp transition between formations of the two types of low-density regions.

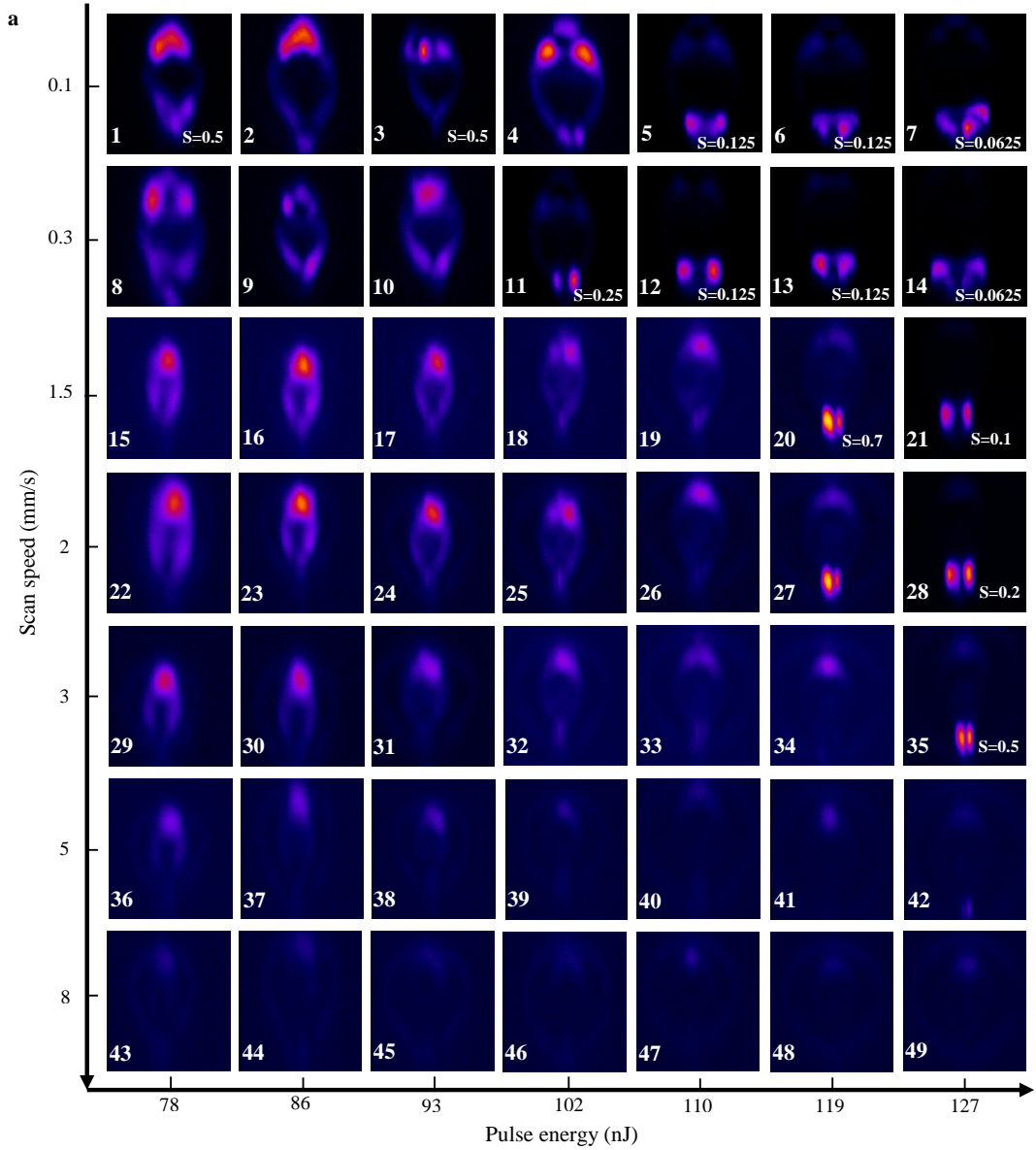

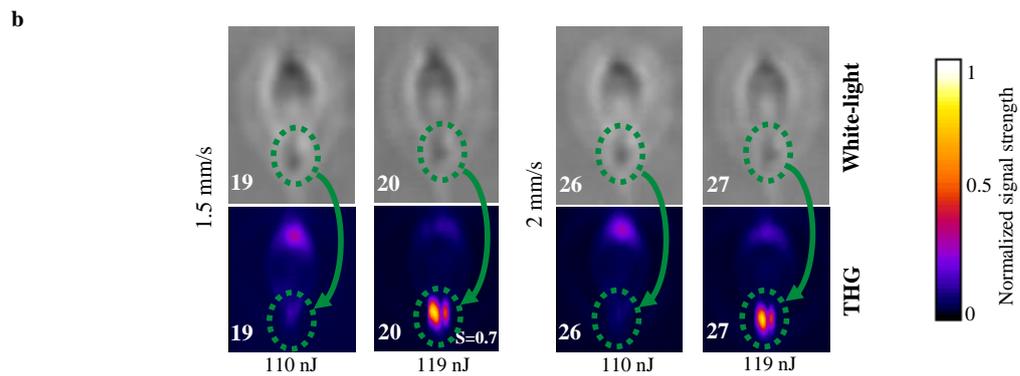

**Figure 1 | Cross-sectional THG images of waveguides in borosilicate glass (in false colour) and comparison between white-light microscopy images and their corresponding THG images for waveguide 19-20 and 26-27.**

**a,** Cross-sectional THG images of waveguides written in borosilicate glass with different pulse energies and scan speeds; the serial number of each waveguide is on the bottom left corner of its THG image; S is the relative integration time of the adaptive THG microscope detector during probing, its default value is 1; S is marked on the bottom right corner of corresponding THG images when it was set to be below 1 to avoid image saturation; some of the THG images were scaled differently to aid visualisation. **b,** Comparison between white-light microscopy images and their corresponding THG images for waveguide 19-20 and 26-27.

The adaptive THG microscope also enabled us to study the micro-morphology of a PIC component from structural and dimensional aspects. To reveal the structural and geometrical profiles of the written waveguides, the THG images shown in Fig. 1a were re-rendered with a different colourmap (See Supplementary Section B). The structural evolution of the waveguides is summarised in Fig. 2a, from which we can see that the waveguides written with moderate scan speeds and pulse energies exhibited clearly what has been referred to elsewhere as "core-cladding" structures[25]. The cladding gradually disappeared as scan speed and pulse energy were decreased but the core tended to disappear with increasing scan speed and pulse energy. Type II low-density regions emerged at high pulse energies and moderate scan speeds. This structural evolution also cannot be fully explained with the existing theories[25, 26]. In Fig. 2b the differences between the measured waveguide outer dimensions in X direction ($OD_X$) and those of waveguides written with scan speed of 0.1 mm/s ($OD_X(0.1)$) are presented; this is to reveal another counterintuitive phenomenon: at pulse energy of 78 and 86 nJ and scan speed of 0.1 and 0.3 mm/s, which are located at the bottom left corner in Fig. 2a with core-only structures, the written waveguides have smaller outer dimensions in X than those written with higher scan speed; once again this phenomenon is inexplicable with the existing theories[25, 26]. However, we found out that this phenomenon correlated with the dimensional evolution of the plasma emission spot generated during waveguide writing. The measured (see Methods) normalized plasma dimension (NPD) result is shown in Fig. 2c; similarly this result is presented in form of NPD – NPD(0.1), where NPD(0.1) is the NPD at scan speed of 0.1 mm/s. From Fig. 2b and c, we can see that there is clear correlation between the outer dimensions in the transverse X direction of written structures and those of the emission plasma spots during writing. Based on this correlation together with previous report on plasma dynamics[29] and distribution controlling[30] during writing, we believe that plasma plays at least an indicative role in the formation of laser written structures.

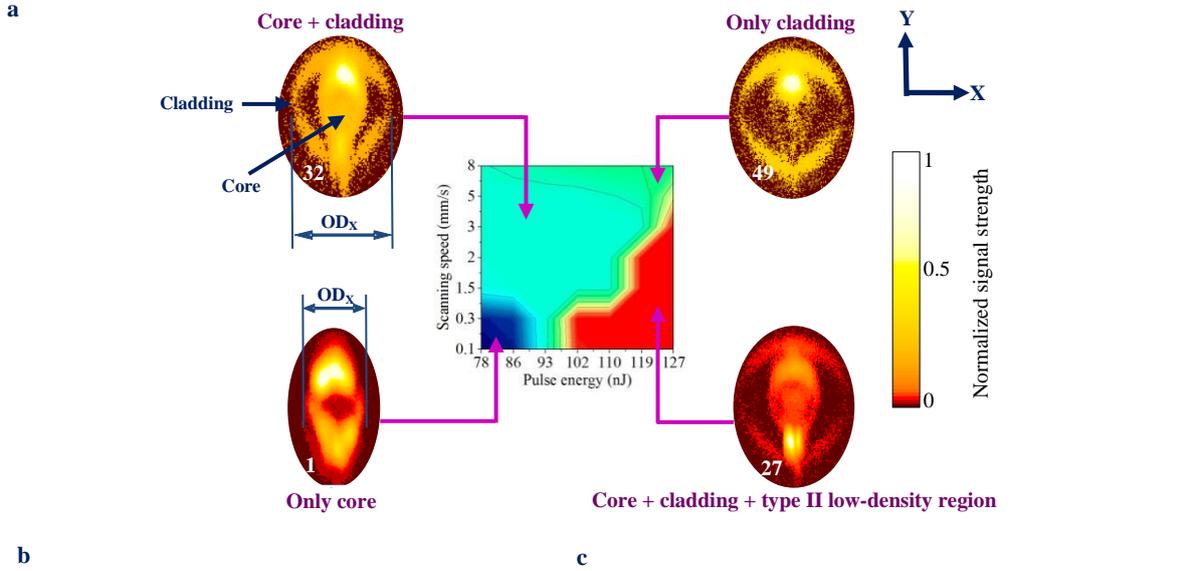

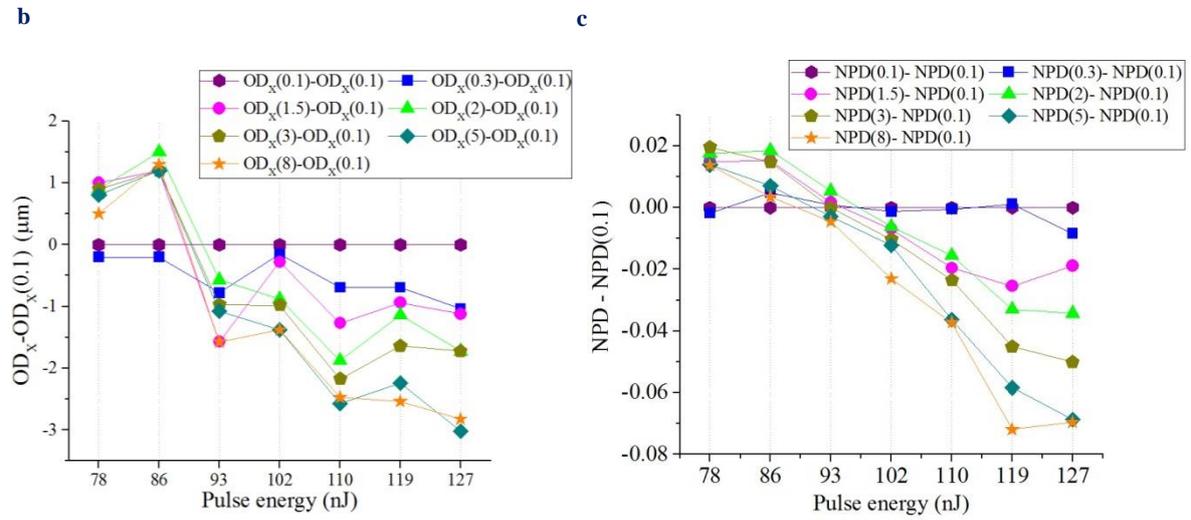

**Figure 2 | Structural and dimensional evolution of waveguides in Fig. 1a and the dimensional evolution of plasma emission spots generated during writing of those waveguides. a,** Under different combinations of writing pulse energy and scan speed, the waveguide cross-sections evolve within four distinct types of structures as exemplified by the four waveguide THG images; each structure type is represented by a colour in the coloured evolution graph; here the THG images were re-rendered with another colourmap to reveal dimensional and structural details. **b,** Difference between the outer dimensions in X ($OD_X$) of all waveguides in Fig. 1a and those of waveguides written with scan speed of 0.1 mm/s ($OD_X(0.1)$); in the legend, the numbers in parentheses are scan speed in mm/s; for complete measured dimensional data please refer to Supplementary Section C. **c,** Difference between NPD and NPD(0.1); NPD is used to represent the normalised dimension of the plasma emission spot generated during writing a waveguide; NPD(0.1) is the NPD at scan speed of 0.1 mm/s; in the legend, the numbers in parentheses are scan speed in mm/s.

**Determining loss, mode and comparative polarization properties of waveguides from THG images**

The performances of those waveguides in terms of propagation loss and near field mode profile were measured (see Methods). From the measurement result shown in Fig. 3 , we can see that the single-mode waveguides (at 788 nm) written with scan speed of 0.3 mm/s exhibited clearly higher propagation losses than those written with higher scan speed. This was indicated by their THG images: as shown in Fig. 1a, THG profiles of waveguides written with scan speed of 0.3 mm/s were micro-morphologically more complex than those written with higher scan speeds. Since more complex micro-morphological THG image or change in the refractive index profile means stronger scattering[31, 32] within waveguide and consequently higher propagation loss. Obviously rounder waveguide shape results in a rounder near field mode profile: as an example shown in the inset of Fig. 3, the near field mode profile of waveguide 8 was rounder than that of waveguide 43, this was because that the cross-sectional shape of waveguide 8 was rounder than that of waveguide 43, as indicated by the measured dimensional data in Supplementary Section C. Therefore, the propagation loss and near field mode profile of a waveguide can be qualitatively assessed with its THG image.

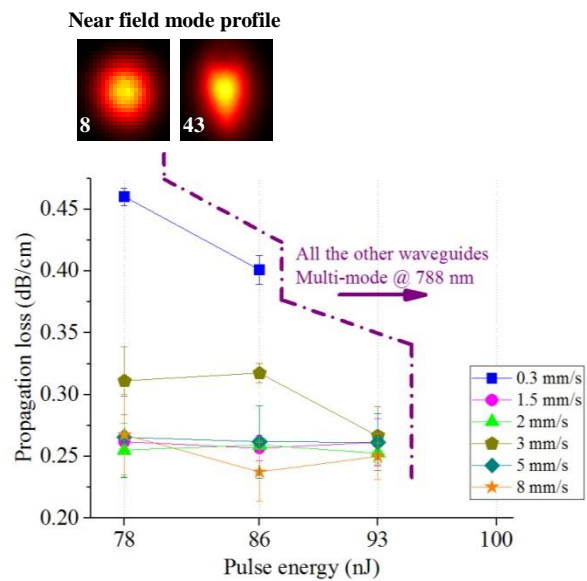

**Figure 3 | Measured propagation losses and near field mode profiles of part of the waveguides shown in Fig 1a.** Measured propagation losses of waveguides that are single mode at wavelength of 788 nm; the error bars represent peak-to-valley variation of multiple tests; insets are the near field mode profiles (in false colour) of waveguide 8 and 43.

The nonreciprocal laser writing phenomenon[33-37] – where waveguides written with laser translation in opposite directions exhibit different properties – was previously discovered only by the difference in appearance under a white-light microscope. However, within certain process window, we found out that although under a white-light microscope there was no obvious difference between waveguides written with only opposite directions, their polarization and loss properties, as well as their THG images, were still different; and the difference between the THG images was more pronounced and distinct than the difference between their polarization or loss properties. We first present here this discovery and then use it as an example to demonstrate how to qualitatively and comparatively determine the polarisation properties of a waveguide relative to another one of the same length, through their THG images.

Three pairs of waveguides were written in fused silica with different pulse energies (see Methods); within each pair, only the scan directions were reversed; other writing parameters and waveguide lengths were kept unchanged. Their cross-sectional THG images are shown in Fig. 4a, where we can see that for the pair written with pulse energy of 23.6 nJ, both the shapes and signal strengths of their THG images were clearly different, which contrasted with the similarity of their white light microscopy images; at pulse energy of 25.2 nJ, their signal strengths were still different but their shapes were similar; as the pulse energy was increased to 26.8 nJ, both their signal strengths and shapes became more similar. The difference in THG image between two waveguides within a pair correlated with the difference between their measured polarization properties and propagation losses in two orthogonal polarization directions (vertical and horizontal) (see Methods); these two differences also exhibited the same trend with increase in writing pulse energy, as shown in Fig. 4. The measured polarization property of each waveguide is shown above its THG image in Fig. 4a, in which the polarization property of a waveguide is represented by the difference between polarization-state ellipse of the input light ($PS_{input}$) and that of output light ($PS_{output}$); $PS_{input}$ was kept unchanged during measurements of all waveguides. Polarization-dependent losses are shown in Fig. 4b. The polarization property of a waveguide includes its birefringence and polarization dependent loss, both of which can be assessed with the THG image as elaborated in Supplementary Section D. So as shown in Fig. 4a, the difference between polarization properties of two waveguides of the same length can be comparatively assessed with their THG images.

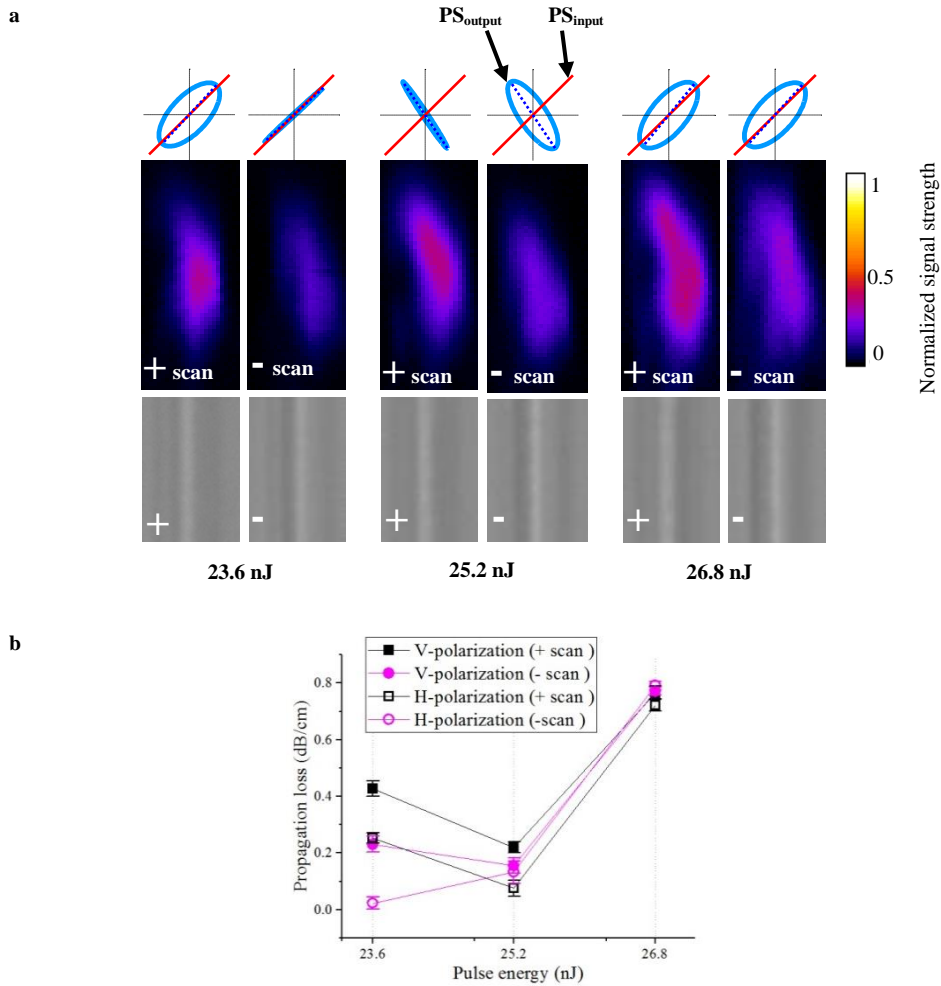

**Figure 4 | Comparative determination of the polarization properties of waveguides from their THG images. a,** Three pairs of waveguides of same length in fused silica. Within each pair, the two waveguides were written with the same writing parameters except in opposite scan directions (represented by '+' and '-'); the middle row figures are the THG images in false colour; on top of each THG image is the measured output polarization state (PS$_{output}$, represented by polarization state ellipse) with respect to that of input (PS$_{input}$); below each THG image is the corresponding white light microscopy image. **b,** Measured propagation losses of the three pairs of waveguides in vertical (V) and horizontal (H) polarization directions; the error bars represent peak-to-valley variation of multiple tests.

**Prediction of complex PIC components' properties based on THG images**

Applications of PICs like photon-based QIPs use complex waveguiding components such as directional couplers and multimode couplers like 3D tritters[16], like those shown schematically in Fig. 5a and b. As examples of how to determine component performance based on THG images, a set of directional couplers and 3D tritters were fabricated using the same selected writing parameters based on the previous study results shown in Fig. 1 and Fig. 3 (See Methods). A simulation method was developed to predict the performance of the directional couplers and the tritters based on solving the coupled mode equations of waveguide optics[16, 38] with information from experimental measurements. This information included the waveguide shape and relative position from THG images and the measured projected refractive index profile[39, 32] (See Supplementary Section E). The simulated output splitting ratios of the directional couplers and experimental measurements are shown in Fig. 5e, from which we can see that the two sets of results agreed well. As shown in the inset of Fig. 5f, the tritter was designed such that three waveguides were situated at the vertices of an isosceles triangle in the coupling region; the triangle was formed by shifting the top waveguide downwards $\Delta Y$ from an equilateral triangle. According to the tritter design, the three waveguides in coupling regions should be equidistant in the X direction, which meant the ratio of $X_{12}/X_{13}$ in Fig. 5g should ideally be unitary; but after fabrication and analysis of the THG images, it was found out that for all the fabricated tritters the ratio of $X_{12}/X_{13}$ was around 0.8, which we believed was due to the nonreciprocal laser writing effect since within each tritter waveguide W1 and W2 were fabricated with scan direction opposite to that of waveguide W3. This slight discrepancy between the designed and measured $X_{12}/X_{13}$ ratios further underlined the importance of THG imaging in PIC fabrication and testing. The simulated output-splitting ratios of the tritters based on the designed and measured $X_{12}/X_{13}$ ratios were both compared with the measured results as shown in Fig. 5f and Fig. 5h. It can clearly be seen that the simulated transmission based on the measured $X_{12}/X_{13}$ ratio agreed well with the measured transmission.

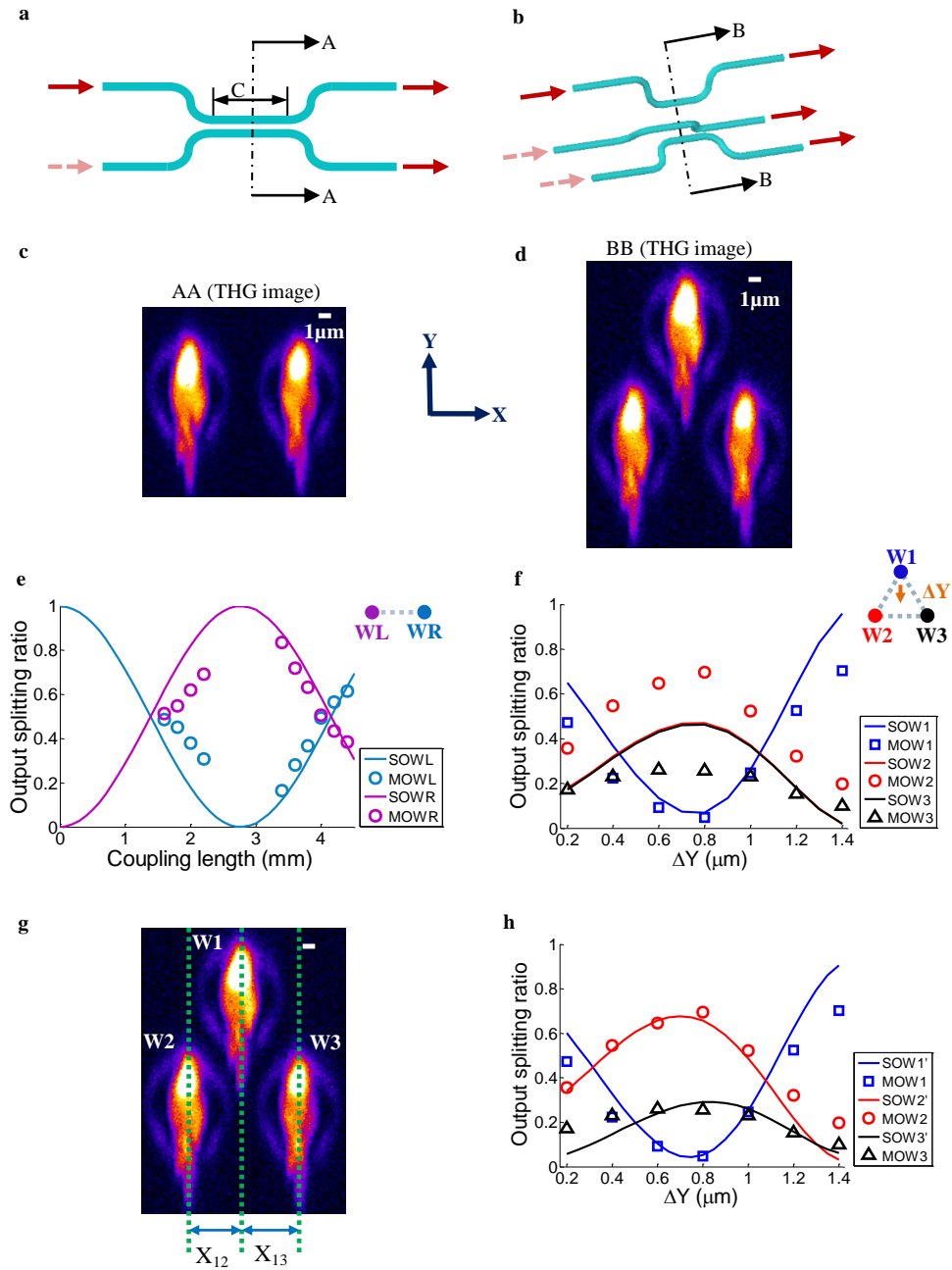

**Figure 5 | Quantitative performance of complex PIC components based on their THG images. a,** Schematic of a directional coupler; C is used to denote the coupling length at the coupling region. **b,** Schematic of a 3D tritter. **c,** Cross-sectional THG image (in false colour) of a directional coupler in its coupling region. **d,** Cross-sectional THG image of a 3D tritter in its coupling region, ΔY was 0.2 μm for this tritter. **e,** Measured and simulated output splitting ratios of a set of directional couplers; the coupling lengths varied from 1.6 mm to 4.4 mm aiming for 50% and 41% reflectivity; in the inset is the schematic of the two waveguides in coupling region, WL and WR denote left and right waveguide respectively; in the legend, SOWL and SOWR are simulated output splitting ratio of waveguide WL and WR respectively, MOWL and MOWR are the measured values; light was coupled into waveguide WL during measurement. **f,** Measured splitting ratios of the tritters and their simulated results based on designed relative position between the three waveguides in coupling regions; the set of tritters were fabricated with ΔY varying from 0.2 mm to

1.4 mm; in the legend, SOW1, 2 and 3 are simulated output splitting ratios of waveguide 1, 2 and 3 respectively; MOW 1, 2 and 3 are their measured values. **g,** The as-built relative position in X-direction between three waveguide in the coupling region of a tritter; $X_{12}$ is the distance between W1 and W2, $X_{13}$ is that between W1 and W3. **g,** Measured splitting ratios of the tritters and their simulated results based on as-built $X_{12}/X_{13}$ from THG images; SOW1, 2' and 3'denote the simulated results based on as-built $X_{12}/X_{13}$ from the THG images.

We would like to point out that all the directional couplers and tritters we have made were tested to be polarization sensitive, especially the tritters; the measured output splitting ratios shown in Fig. 5e, f and h were taken with a vertical (Y axis in Fig. 5 ) polarized input beam (See Methods). As detailed in Supplementary Section E, for computational efficiency our current simulation methods are based on solving the coupled mode equations of waveguide optics[16, 38] with assumption of identical and step-function refractive index change profile for all the waveguides in a PIC component. This means that our current simulation is able to perform only polarization-independent simulations. However we believe this limitation could be removed by performing the simulation completely in a finite element numerical solver like COMSOL, especially when the complete 3D THG image of a PIC component, instead of only a cross-sectional THG image, is inputted into COMSOL for simulation.

This demonstration shows clearly that simulations of the performance of complex PIC components like the directional coupler and the tritter can be improved using information obtained using THG microscopy.

**Conclusion and outlook**

Based on a custom-built adaptive THG microscope we have demonstrated a method to test individual components inside a PIC by assessing their performance based on their THG images; this method has been demonstrated for both basic quantum PIC component and more complex components. This component-wise testing method will enable the further scaling up of PICs, which is currently bottlenecked by fabrication errors and variations in substrate uniformity. It is likely that fabrication becomes even more challenging, given the ever-stricter requirements on and fast-growing complexity of PIC demanded by many applications like photon-based quantum computing. Moreover, we also demonstrated that this method provides a new pathway for us to study the fundamentals of light-matter interaction, through discovering following new phenomena as examples: I) the micro-morphological evolution of waveguides along writing parameters, including the evolution of core-cladding structure and two types of low-density region; II) the correlation between the waveguide cross-sectional dimension and the dimension of plasma emission spot during writing.

Although this method was demonstrated with femtosecond-laser-written PICs in glasses, we expect this method can be directly extended to component-wise testing any silica-on-silicon PIC like UV-written[19, 40, 41] and CMOS-process-based[21, 42-45] ones, since the wavelength of the THG microscope excitation laser is within the transmission window of silicon.

**Methods**

**Femtosecond laser writing.** The laser, which was employed to write the integrated photonic components inside glass, was the second harmonic of a regenerative amplified Yb:KGW laser (Light Conversion Pharos SP-06-1000-pp) with 1 MHz repetition rate, 514 nm wavelength, 170 fs pulse duration laser. The power of the circularly-polarized laser beam was regulated through combination of a motorised rotating half waveplate and a polarization beam splitter before being focused 120 μm below the top surface of the glass chip with a 0.5 NA objective lens. The glass chip, which was fixed on a three-axis air bearing stage (AerotechABL10100L/ABL10100L/ANT95-3-V), was transversely scanned relative to the focus to inscribe integrated photonic components. The waveguides were written in borosilicate glass (Corning EAGLE 2000), with scan speeds from 0.1 mm/s to 8 mm/s and pulse energies from 78 nJ to 127 nJ, measured before the sample surface; in fused silica (LEONI SQ0), all the integrated photonic components were written with a scan speed of 10 µm/s and pulse energies from 23.6 nJ to 26.8 nJ before the sample surface.

All the directional couplers and 3D tritters were written with pulse energy of 85nJ and scanning speed of 2 mm/s, in Corning EAGLE 2000 glass; the writing parameters were chosen to balance between propagation loss and mode shape. The radius of all the curved parts of the directional couplers and 3D tritters was 50 mm.

The separation between two waveguides in the coupling regions of all written directional couplers was 8 µm. Every 3D tritter was designed in a way that the three waveguides were situated at vertices of an isosceles triangle in the coupling region; the triangle was formed by shifting the top waveguide downwards ΔY from an equilateral triangle with side length of 8 µm. The set of 3D tritters were fabricated with ΔY varying from 0.2 µm to 1.4 µm.

**THG imaging.** The excitation laser of the custom-built adaptive THG microscope was a chromium forsterite laser with pulse duration of 65 fs, repetition rate of 100 MHz and centre wavelength of 1235 nm; after being expanded, the laser beam was scanned by a two-axis galvanometer scanner; A deformable membrane mirror (MIRAO 52-e, Imagine Eyes) was employed to correct the optical aberration introduced by the microscope itself and the specimens; then the beam was focused by water immersion objective lens (Olympus UApo/340,

NA = 1.15) to a laser written PIC which was held by a piezo stage on top of a two-dimensional manual stage. The generated third-harmonic light was collected by a condenser before being filtered and detected by a photomultiplier tube (PMT). Three-dimensional scanning of the specimen was realized through a galvanometer scanner (X and Y) and a piezo stage (Z). The aberration correction procedure involved sequentially adjusting the amplitudes of the Zernike polynomial modes added to the deformable mirror, in order to maximize an image quality metric, which was defined as the total image intensity. First, system aberrations resulted from the microscope light path were corrected by optimizing the THG signal from the rear surface of a cover glass. Then the remaining aberrations, which were introduced by the chip substrate and the depth of the component to be probed, were corrected (Supplementary Section A).

**Measurement of the cross-sectional dimension of a waveguide.** All the cross-sectional dimensions were determined through counting THG image pixels whose pixel-to-dimension ratio has been calibrated through features with known dimensions. For those waveguides with very strong type II low-density regions (corresponding to very strong THG signal), the cladding regions and the type II low-density regions were imaged separately to guarantee the visibilities of the cladding regions (as an example, separate THG images of waveguide 14 were shown in Fig. S3 in Supplementary Section B).

**Measurement of the dimension of plasma emission spot during laser writing.** During waveguides writing, the generated plasma emission spots were recorded with an in-line camera (Baumer VLU-12M). For repeatability of this measurement, the dimension of the plasma emission spot was determined in following way: the camera settings, which were chosen to guarantee that there was always saturation only in the plasma emission spot, were fixed throughout fabrication of all waveguides; in this way, the measured dimension corresponded to an intensity threshold of plasma emission, bearing in mind that there is also an intensity threshold of writing laser for refractive index change. It is worth mentioning that during this measurement all the waveguides were written in the same depth. The normalized dimension of plasma spot $ND_p$ was defined as

$$ND_p = \frac{\sqrt{aos}}{\sqrt{aos_{max}}}$$

where $aos$ is area of saturation region of the recorded plasma emission spot; $aos$ is in number of saturated CCD pixels. And $aos_{max}$ is the maximum value among all $aos$.

**Performance of laser written components.** All the propagation losses were measured through the cutback approach. For waveguides in borosilicate glass (Corning EAGLE 2000), a 788 nm wavelength (measured with an Ocean Optics USB2000 spectrometer) laser beam was coupled into each waveguide through butt-coupling with a 20-meter-long single mode fiber (Thorlabs S630-HP); the output near field mode profile of a waveguide was recorded with a CCD camera through an objective lens and a tube lens. A photodiode power sensor was used to measure the power of the output beam from a waveguide. To further reduce the almost negligible influence of the unguided light after the objective lens, the power sensor was placed 220 mm behind the objective lens and an iris was inserted before the power sensor when the output power was measured. When the polarization sensitive waveguides in fused silica (LEONI SQ0) were tested, the laser source and butt-coupling fibre were changed to a 777 nm wavelength linearly polarized laser and a polarization-maintaining fibre (Thorlabs P1-630PM-FC-5) respectively.

The polarization states of the input and output beam from a waveguide in fused silica were measured with a polarimeter (Polarization Analyzer SK010PA-VIS, Schäfter + Kirchhoff GmbH)

The performances of the directional couplers and 3D tritters were measured through butt-coupling with polarization-maintaining fibre from the 777 nm wavelength laser (Thorlabs P1-630PM-FC-5) at vertical polarization.

**Acknowledgments**

We are grateful to Ian Walmsley for constructive discussions on this paper. This work was support by the UK Engineering and Physical Sciences Research Council through grants EP/M013243/1 and EP/K034480/1. This publication was supported by the Oxford RCUK Open Access Block Grant and accordingly underlying research materials are available upon request.

**Author contributions**

J.G. conceived the idea, coordinated the project, carried out all the research works except THG imaging, refractive index measurement and directional coupler and tritter simulations, and drafted the manuscript. X.L. performed all the THG imaging and refractive index measurement. A. J. M. developed the simulation method for directional coupler and tritter simulation. M. J. B. oversaw this project and contributed to problem solving. All authors contributed to the manuscript writing.

**Competing interests**

The authors declare no competing financial interests.

**References**

1. Sansoni, L. *et al.* Polarization entangled state measurement on a chip. *Phys. Rev. Lett.* **105**, 200503 (2010).

2. Crespi, A. *et al.* Integrated photonic quantum gates for polarization qubits. *Nat. Commun.* **2**, 566 (2011).

3. Crespi, A. *et al.* Anderson localization of entangled photons in an integrated quantum walk. *Nat. Photonics* **7**, 322-328 (2013).

4. Di Giuseppe, G. *et al.* Einstein-podolsky-Rosen spatial entanglement in Ordered and Anderson photonic lattices. *Phys. Rev. Lett.* **110**, 150503 (2013).

5. Tillmann, M. *et al.* Experimental boson sampling. *Nat. Photonics* **7**, 540-544 (2013).

6. Chaboyer, Z. *et al.* Tunable quantum interference in a 3D integrated circuit. *Sci. Rep.* **5**, 9601 (2015).

7. Meany, T. *et al.* Engineering integrated photonics for heralded quantum gates. *Sci. Rep.* **6**, 25126 (2016).

8. Ciampini, M. A. *et al.* Path-polarization hyperentangled and cluster states of photons on a chip. *Light Sci. Appl.* **5**, e16064 (2016).

9. Cheng, Y. Sugioka, K. & Midorikawa, K. Microfluidic laser embedded in glass by three-dimensional femtosecond laser microprocessing. *Opt. Lett.* **29**, 2007-2009 (2004).

10. Thomson, R. R., Birks, T. A., Leon-Saval, S. G., Kar, A. K. & Bland-Hawthorn, J. Ultrafast laser inscription of an integrated photonic lantern. *Opt. Express* **19**, 5698-5705 (2011).

11. Sun, B., Salter, P. S. & Booth, M. J. High conductivity micro-wires in diamond following arbitrary paths. *Appl. Phys. Lett.* **105**, 231105 (2014).

12. Douglass, G., Dreisow, F., Gross, S., Nolte, S. & Withford, M. J. Towards femtosecond laser written arrayed waveguide gratings. *Opt. Express* **23**, 21392-21402 (2015).

13. Nie, W. *et al.* Room-temperature subnanosecond waveguide lasers in Nd: $YVO_4$ Q-switched by phase-change $VO_2$: A comparison with 2D materials. *Sci. Rep.* **7**, 46162 (2017).

14. Buch-Månson, N. *et al.* Rapid prototyping of polymeric nanopillars by 3D direct laser writing for controlling cell behavior. *Sci. Rep.* **7**, 9247 (2017).

15. Rechtsman, M. C. *et al.* Photonic floquet topological insulators. *Nature* **496**, 196-200 (2013).

16. Spagnolo, N. *et al.* Three-photon bosonic coalescence in an integrated tritter. *Nat. Commun.* **4**, 1606 (2013).

17. Sansoni, L. *et al.* Two-particle Bosonic-Fermionic quantum walk via integrated photonics. *Phys. Rev. Lett.* **108**, 010502 (2012).

18. Poulios, K. *et al.* Quantum walks of correlated photon pairs in two-dimensional waveguide arrays. *Phys. Rev. Lett.* **112**, 143604 (2014).

19. Metcalf, B. J. *et al.* Quantum teleportation on a photonic chip *Nat. Photonics* **8**, 770-774 (2014).

20. Flamini, F. *et al.* Thermally reconfigurable quantum photonic circuits at telecom wavelength by femtosecond laser micromachining. *Light Sci. Appl.* **4**, e354 (2015).

21. Dai, D-X., Bauters, J. & Bowers, J. E. Passive technologies for future large-scale photonic integrated circuits on silicon: polarization handling, light non-reciprocity and loss reduction. *Light Sci. Appl.* **1**, e1 (2012).

22. Bruck, R. *et al.* Device-level characterization of the flow of light in integrated photonic circuits using ultrafast photomodulation spectroscopy. *Nat. Photonics* **9**, 54-60 (2014).

23. Müller, M., Squier, J., Wilsion, K. R. & Brakenhoff, G. J. 3D microscopy of transparent objects using third-harmonic generation. *J. Microsc.* **191**, 266-274 (1998).

24. Marshall, G. D., Jesacher, A., Thayil, A., Withford, M. J. & Booth, M. Three-dimensional imaging of direct-written photonic structures. *Opt. Lett*. **36**, 695-697 (2011).

25. Eaton, S. M. *et al.* Transition from thermal diffusion to heat accumulation in high repetition rate femtosecond laser writing of buried optical waveguides. *Opt. Express* **16**, 9443-9458 (2008).

26. Fernandez, T. T. Ion migration assisted inscription of high refractive index contrast waveguides by femtosecond laser pulses in phosphate glass. *Opt. Lett*. **38**, 5248-5251 (2013).

27. Tan, D-Z., Sharafudeen, K. N., Yue, Y-Z. & Qiu, J-R. Femtosecond laser induced phenomena in transparent solid materials: Fundamentals and applications. *Prog. Mater. Sci.* **76**, 154-228 (2016).

28. Kanehira, S., Si, J-H., Qiu, J-R., Fujita, K. & Hirao, K. Periodic nanovoid structures via femtosecond laser irradiation. *Nano Lett*. **5**, 1591-1595 (2005).

29. Miyamoto, I. *et al.* Mechanism of dynamic plasma motion in internal modification of glass by fs-laser pulses at high pulse repetition rate. *Opt. Express* **24**, 25718-25731 (2016).

30. Fernandez, T. T. *et al.* Controlling plasma distributions as driving forces for ion migration during fs laser writing. *J. Phys. D: Appl. Phys.* **48**, 155101 (2015).

31. Eaton, S. M. Spectral loss characterization of femtosecond laser written waveguides in glass with application to demultiplexing of 1300 and 1550 nm wavelengths. *J. Lightwave Technol.* **27**, 1079–1085 (2009).

32. Guan, J., Liu, X., Salter, P. S., & Booth, M. J. Hybrid laser written waveguides in fused silica for low loss and polarization independence. *Opt. Express* **25**, 4845-4859 (2017).

33. Kanzansky, P. G. *et al.* "Quill" writing with ultrashort light pulses in transparent materials. *Appl. Phys. Lett*. **90**, 151120 (2007).

34. Yang, W-J., Kazansky, P. G. & Svirko, Y. P. Non-reciprocal ultrafast laser writing. *Nat. Photonics* **2**, 99-104 (2008).

35. Vitek, D. N. *et al.* Spatio-temporally focused femtosecond laser pulses for nonreciprocal writing in optically transparent materials. *Opt. Express* **18**, 24673-24678 (2010).

36. Salter, P. S. & Booth, M. J. Dynamic control of directional asymmetry observed in ultrafast laser direct writing. *Appl. Phys. Lett*. **101**, 141109 (2012).

37. Patel, A., Svirko, Y., Durfee, C. & Kazansky, P., G. Direct writing with tilted –front femtosecond pulses. *Sci. Rep.* **7**, 12928 (2017).

38. Suzuki, K., Sharma, V., Fujimoto, J. G., Ippen, E. P. and Nasu, Y. Characterization of symmetric [3 × 3] directional couplers fabricated by direct writing with a femtosecond laser oscillator. *Opt. Express* **14**, 2335-2343 (2006).

39. Jesacher, A., Salter, P. S. and Booth, M. J. Refractive index profiling of direct laser written waveguides: tomographic phase imagine. *Opt. Mater. Express* **3**, 1223-1232 (2013).

40. Spring, J. B. et al. Boson sampling on a photonic chip. *Science* 339, 798-801 (2013).

41. Metcalf, B. J. et al. Multiphoton quantum interference in a multiport integrated photonic device. *Nat. Commun*. **4**, 1356 (2013).

42. Politi, A., Cryan, M. J., Rarity, J. G., Yu, S-Y. & O'Brien, J. L. Silica-on-silicon waveguide quantum circuits. *Science* **320**, 646-649 (2008).

43. Silverstone, J. W. *et al.* Qubit entanglement between ring-resonator photon-pair sources on a silicon chip. *Nat. Commun*. **6**, 7948 (2015).

44. Carolar, J. *et al.* Universal linear optics. *Science* **349**, 711-716 (2015).

45. Feng, L-T. *et al.* On-chip coherent conversion of photonic quantum entanglement between different degrees of freedom. *Nat. Commun*. **7**, 11985 (2016).